%% file: matching.tex
\documentclass[%
a4paper,%
USenglish,%
thm-restate,%
]{lipics-v2021}

\title{
On $b$-Matching and Fully-Dynamic Maximum $k$-Edge Coloring
}

\author{Antoine El-Hayek}{%
Institute of Science and Technology Austria (ISTA), Klosterneuburg, Austria}{%
antoine.el-hayek@ist.ac.at}{%
https://orcid.org/0000-0003-4268-7368
}{%
}

\author{Kathrin Hanauer}{%
Faculty of Computer Science, University of Vienna, Austria}{%
kathrin.hanauer@univie.ac.at}{%
https://orcid.org/0000-0002-5945-837X
}{}

\author{Monika Henzinger}{%
Institute of Science and Technology Austria (ISTA), Klosterneuburg, Austria}{%
monika.henzinger@ist.ac.at}{%
https://orcid.org/0000-0002-5008-6530
}{}

\authorrunning{A. El-Hayek and K. Hanauer and M. Henzinger} %

\Copyright{Antoine El-Hayek and Kathrin Hanauer and Monika Henzinger} %



\category{} %

\relatedversion{} %

\funding{This project has received funding from the European Research Council (ERC) under the European Union's Horizon 2020 research and innovation programme (MoDynStruct, No. 101019564)  \includegraphics[width=0.9cm]{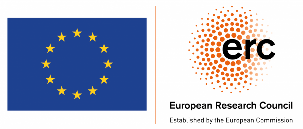} and the Austrian Science Fund (FWF) grant  \href{https://www.doi.org/10.55776/Z422}{DOI 10.55776/Z422}, grant  \href{https://www.doi.org/10.55776/I5982}{DOI 10.55776/I5982}, and grant  \href{https://www.doi.org/10.55776/P33775}{DOI 10.55776/P33775} with additional funding from the netidee SCIENCE Stiftung, 2020–2024.
This work was further supported by the Federal Ministry of Education and Research (BMBF) project, 6G-RIC: 6G Research and Innovation Cluster, grant 16KISK020K.}

\nolinenumbers %

\usepackage{graphicx}

\usepackage{amsfonts}
\usepackage[bbgreekl]{mathbbol}
\DeclareSymbolFontAlphabet{\mathbb}{AMSb}
\DeclareSymbolFontAlphabet{\mathbbl}{bbold}

\usepackage{caption}
\usepackage{subcaption} 
\usepackage{mathtools}
\usepackage{mathbbol}
\usepackage{soul} %
\usepackage[ruled]{algorithm2e}
\usepackage{booktabs}

\SetKwBlock{Begin}{In round $k: 1 \leq k \leq \broadcasttime$ do}{end}
\SetKwInput{KwInput}{Input}
\SetKwInput{KwOutput}{Output}

\usepackage{array}

\usepackage{amsmath}
\usepackage{graphicx}
\usepackage{graphics}
\usepackage{amssymb}
\usepackage{amsthm}
\usepackage{thmtools}
\usepackage{hyperref}

\newcommand{\ceil}[1]{\left\lceil #1 \right\rceil}
\newcommand{\floor}[1]{\left\lfloor #1 \right\rfloor}
\newcommand{\union}{\cup}

\newcommand{\Union}{\bigcup}

\newcommand{\card}[1]{\left|#1\right|}
\newcommand{\N}{\mathbb{N}}

\newcommand{\Pea}{\mathcal{P}}
\newcommand{\Proba}{\mathbb{P}}
\newcommand{\expect}{\mathbb{E}}

\newcommand{\Alg}{\mathcal{A}}
\newcommand{\Var}{\mathrm{Var}}
\newcommand{\1}{\mathbbl{1}}
\newcommand{\bvec}[1]{\mathbf{#1}}
\renewcommand{\texorpdfstring}[1]{}
\DeclareMathOperator{\polylog}{polylog}
\DeclareMathOperator{\poly}{poly}
\newcommand{\Restrict}[2]{#1\big|_{#2}}

\newcommand{\Greedy}{\texttt{Greedy}}
\newcommand{\MatchO}{\texttt{MatchO}}
\newcommand{\MatchA}{\texttt{MatchA}}
\newcommand{\bMatchO}{\texttt{bipartite MatchO}}
\newcommand{\bMatchA}{\texttt{bipartite MatchA}}

\newcommand{\MkEC}{\textsc{MkEC}}
\newcommand{\etal}{et~al.}

\EventEditors{Kitty Meeks and Christian Scheideler}
\EventNoEds{2}
\EventLongTitle{4th Symposium on Algorithmic Foundations of Dynamic Networks (SAND 2025)}
\EventShortTitle{SAND 2025}
\EventAcronym{SAND}
\EventYear{2025}
\EventDate{June 9--11, 2025}
\EventLocation{Liverpool, GB}
\EventLogo{}
\SeriesVolume{330}
\ArticleNo{4}

\usepackage[colorinlistoftodos,prependcaption,textsize=tiny]{todonotes}
\newcommand{\antoine}[1]{\todo[linecolor=orange,backgroundcolor=orange!25,bordercolor=orange]{{\bf Antoine:} #1}}
\newcommand{\mh}[1]{\todo[linecolor=blue,backgroundcolor=blue!25,bordercolor=blue]{{\bf Monika:} #1}}
\newcommand{\kh}[1]{\todo[linecolor=purple,backgroundcolor=purple!25,bordercolor=purple]{{\bf Kathrin:} #1}}

\newcommand{\oldtext}[1]{{}}
\renewcommand{\antoine}[1]{}
\renewcommand{\mh}[1]{}
\renewcommand{\kh}[1]{}

\date{}

\begin{document}
\maketitle \sloppy
\input{1-abstract}
\input{2-introduction}
\input{222-related-work-short}
\input{11-Technical_Overview}
\input{10-preliminaries}

\input{5-Matching_polytope}

\input{7-sparsificationscheme}

\input{8-sparsificationlemma}

\input{4-Dynamic}

\input{9-conclusion}

\bibliography{matchingbib}
\newpage
\appendix
\input{appendix}
\input{22-related-work}

\end{document}

%% file: 1-abstract.tex
\begin{abstract}
Given a graph $G$ that undergoes a sequence of edge insertions and deletions, we study the \textsc{Maximum $k$-Edge Coloring} problem (\MkEC): Having access to $k$ different colors, color as many edges of $G$ as possible such that no two adjacent edges share the same color.
While this problem is different from simply maintaining a $b$-matching with $b=k$, the two problems are related.
However, maximum $b$-matching can be solved efficiently in the static setting, whereas \MkEC{} is NP-hard and even APX-hard for $k \ge 2$. 

We present new results on both problems:
For $b$-matching, we show a new integrality gap result
and we adapt Wajc's matching sparsification scheme~\cite{DBLP:conf/stoc/Wajc20}
for the case where $b$ is a constant.

Using these as basis, we give three new algorithms for the dynamic \MkEC{} problem:
Our \MatchO{} algorithm builds on the dynamic $(2+\epsilon)$-approximation algorithm of Bhattacharya, Gupta, and Mohan~\cite{BhattacharyaGM17} for $b$-matching and achieves a $(2+\epsilon)\frac{k+1} k$-approximation in $O(poly(\log n, \epsilon^{-1}))$ update time against an \emph{oblivious} adversary.
Our \MatchA{} algorithm builds on the dynamic $(7+\epsilon)$-approximation algorithm by Bhattacharya, Henzinger, and Italiano~\cite{DBLP:conf/soda/BhattacharyaHI15} for fractional $b$-matching and achieves a $(7+\epsilon)\frac{3k+3}{3k-1}$-approximation in $O(poly(\log n, \epsilon^{-1}))$ update time against an \emph{adaptive} adversary. Moreover, our reductions 
use the dynamic $b$-matching algorithm as a black box, so any future improvement in the approximation ratio for dynamic $b$-matching will automatically translate into a better approximation ratio for our algorithms.
Finally, we present a greedy algorithm with $O(\Delta+k)$ update time, which guarantees a $2.16$~approximation factor.

\end{abstract}

%% file: 2-introduction.tex
\section{Introduction}

In %
large data centers, new technologies such as optical switches allow for quick adaptations of the network topology that are optimally tailored to current traffic demands.
Indeed, network performance has been identified to be a major bottleneck for the scalability
of computations in the cloud~\cite{khani2021sip,mogul2012we}.
Each optical switch can establish a set of direct connections between pairs of data center racks, such that each rack has a high-bandwidth connection to at most one other rack via the switch.
The regular network infrastructure remains in place and can be used for all other traffic.
With their fast reconfiguration time and the possibility to use multiple appliances in parallel, optical switches have become a promising means to mitigate the performance bottleneck.
A major algorithmic challenge is how to configure them optimally.

Consider an undirected graph $G$ where each node represents a data center rack and the edges indicate pairs of racks with high communication demand.
We refer to $G$ as \emph{demand graph}.
Each set of connections realized by a single optical switch is a \emph{matching} in $G$, i.e., a set of pairwise node-disjoint edges, and it is desirable that this matching is large, such that a substantial amount of the traffic is routed via the high-bandwidth direct connections.
For $k \in \mathbb{N}$ optical switches, the problem hence amounts to finding a collection of $k$ pairwise edge-disjoint matchings $M_i$, $1 \leq i \leq k$, with maximum total cardinality $\sum_{i=1}^k |M_i|$.
We call this the \emph{maximum $k$-disjoint matching problem}.
Identifying each matching with a unique color, it can equivalently be rephrased as a \emph{maximum $k$-edge coloring} problem (\MkEC{}), where the goal is to maximize the number of colored edges.
As communication demands naturally change frequently over time, we study the problem in
the dynamic setting.
In some situations, e.g. when remote resources such as GPUs are accessed via the network, the demand graph may be bipartite, which is why we also consider bipartite graphs.

There exists a related, but different problem, the \emph{maximum $\bvec{b}$-matching
problem}: for a graph $G = (V, E)$ %
and $\bvec{b} \in \mathbb{N}^V$%
, find a maximum-cardinality subset of edges $H \subseteq E$ such that each vertex $v \in V$ is incident to at most $b_v$ edges in $H$.
Note that one can set $b_v$ to a constant $k$ for all $v \in V$, but this does not yield the
\MkEC{} problem:
there is no requirement that $H$ can be edge-colored completely with $k$ colors.
Consider, e.g., a graph $G = (V, E)$ that is a length-3 cycle. 
A solution $H$ to the $\bvec{b}$-matching problem with $b_v=2 \;\forall v \in V$ contains all three edges of $G$, while a solution to the maximum 2-edge coloring problem can only color two edges of $G$.
The third edge has to remain uncolored.
This example shows that an optimal solution to the \MkEC{} problem can be 1.5 times smaller than one to the $\bvec{b}$-matching problem with $b_v = k$ for all $v \in V$.
Furthermore, a solution to the latter does not always give a solution to the former.
In general, deciding whether a graph with maximum degree $\Delta$ can be edge-colored with $\Delta$ colors
or whether $\Delta+1$ are required is a well-known NP-hard problem~\cite{DBLP:journals/siamcomp/Holyer81a} (Vizing~\cite{vizing1964} showed that every simple graph needs either $\Delta$ or $\Delta(G)+1$ colors to color all edges).
On bipartite graphs, however, $\Delta$ colors always suffice.

Let $G = (V, E)$ be a graph with $n = |V|$ and $m = |E|$.
For the fully dynamic $\bvec{b}$-matching problem, Bhattacharya, Henzinger, and Italiano~\cite{primaldual} gave a constant approximation algorithm with $O(\log^3 n)$ update time which works against an adaptive adversary.
If the adversary is oblivious,
there also is a $(2 + \epsilon)$-approximation with 
$O(1/\epsilon^4)$ update time by Bhattacharya, Gupta, and Mohan~\cite{BhattacharyaGM17}.

The only prior work for the fully dynamic \MkEC{} problem is an experimental analysis of various dynamic algorithms by Hanauer, Henzinger, Ost and Schmid~\cite{infocomdynamic},
who, among others, also describe a near-linear-time, fully-dynamic $3$-approximation algorithm for the weighted case.
Dropping the weights, we show how to significantly improve the update time to $O(\poly(\log n, \epsilon^{-1}))$ while achieving an approximation ratio to
almost $(2 + 2/k)$ against an \emph{oblivious} adversary and  $(7+\epsilon)(1+4/(3k-1))$ (for any $\epsilon >0$) against an \emph{adaptive} adversary.
The problem is known to be NP-hard and even APX-hard for $k \geq 2$~\cite{originalpaper}.

\input{table-algorithms}
\textbf{Our contributions.}
We show that the integrality gap for the $\bvec{b}$-matching problem is $\frac {3\beta}{3\beta -1}$ where $\beta = \min_{v \in V} b_v$. In the case where $\bvec{b}$ is the constant vector with $b_v = k\; \forall v \in V$, we adapt the elegant rounding technique given by Wajc~\cite{DBLP:conf/stoc/Wajc20}, who showed how to round a fractional matching to an integral matching in a dynamic graph, to round a given fractional $k$-matching\footnote{Whenever $\bvec{b}$ is the all-$k$ vector for some constant $k$, we will refer to the problem as $k$-matching instead of $\bvec{b}$-matching.} to an integral $k$-matching whose size is linear (up to polylogarithmic factors) in the size of the optimal solution.

For the dynamic \MkEC{} problem, we describe and analyze three dynamic approximation algorithms, with different trade-offs between their update time and approximation ratio. We also give two versions specific to bipartite graphs, where the approximation ratio is improved.
See Table~\ref{tab:algorithms} for a summary. Our algorithms \MatchO{} and \MatchA{} represent a general black-box reduction from dynamic maximum $k$-edge coloring to dynamic $\bvec{b}$-matching. Thus, any improvement in the approximation ratio of maximum $\bvec{b}$-matching immediately leads to an improvement of the approximation ratio of our algorithms.

Most proofs are only available in the appendix due to space restrictions.

%% file: table-algorithms.tex
\begin{table}
\caption{Previous and New Algorithms for Dynamic Maximum $k$-Edge-Coloring. The ``Det.?'' column states whether or not the algorithm is deterministic.}\label{tab:algorithms}
\centering
\addtolength{\tabcolsep}{-2.5pt}
\begin{tabular}{@{}lccccrr@{}}
\toprule
Algorithm & Update Time & Approximation Ratio & Det.?&  Adversary & Theorem & Sect.\\
\midrule
\multirow{2}{*}{algorithms in~\cite{infocomdynamic}} & $O(n)$ or more & $3$ & yes & -- & \\ %
& $O(1)$ & unknown & no & -- & & \\ %
\midrule
\Greedy                 & $O(k + \Delta)$              & $1+2\frac{\sqrt 3}3 \approx 2.155$                              & yes & --  & \ref{thm:greedy} & \ref{sect:greedy-dyn}\\
\MatchO & $O(\poly(\log n, \epsilon^{-1}))$ & $(2+\epsilon)(1+1/k)$                      &no  & oblivious & \ref{thm:matcho} & \ref{sect:matcho}\\
\tikz\draw[thick] (4pt,0) -| (0,4pt); \quad\texttt{bipartite} & $O(\poly(\log n, \epsilon^{-1}))$ & $(2+\epsilon)$ %
&no  & oblivious & \ref{thm:matcho-bip} &\ref{sect:matcho}\\
\MatchA  & $O(\poly(\log n, \epsilon^{-1}))$ & $(7+\epsilon)\frac{3k+3}{3k-1}$            &no & adaptive & \ref{thm:matcha}  & \ref{sect:matcha}\\
\tikz\draw[thick] (4pt,0) -| (0,4pt); \quad\texttt{bipartite} & $O(\poly(\log n, \epsilon^{-1}))$ & $(7+\epsilon)$            &no & adaptive & \ref{thm:matcha-bip}  & \ref{sect:matcha}\\
\bottomrule
\end{tabular}
\end{table}

%% file: 222-related-work-short.tex
\section{Related Work}\label{sect:related-short}
We only give a short overview here, see \autoref{sect:related} for an extended version.

\textbf{Edge Coloring.}
Given a graph $G$ of maximum degree $\Delta$, its \emph{chromatic index} $\chi'(G)$ is the smallest value $q$ such that all edges of $G$ can be colored with $q$ colors.
Generally, $\Delta \leq \chi'(G) \leq \Delta+1$~\cite{vizing1964},
whereas $\chi'(G) = \Delta$~\cite{konig1916graphen} for bipartite graphs.
Deciding whether $\chi'(G) = \Delta$ or $\chi'(G) = \Delta + 1$ is NP-hard already for $\Delta = 3$~\cite{DBLP:journals/siamcomp/Holyer81a}, even if $G$ is regular~\cite{nphard}.

For an $n$-vertex $m$-edge graph $G$,
Gabow~\cite{gabowcolor} gave an $O(m \Delta \log n)$-time coloring algorithm that uses at most
$\Delta+1$ colors.
Misra and Gries~\cite{misragries} gave an algorithm that needs $O(mn)$ running time,
and which was improved to $O(m\sqrt{n})$ running time by 
Sinnamon~\cite{DBLP:journals/corr/abs-1907-03201}. A recent series of works further reduced the running time to $\tilde O (n^2)$ (Assadi~\cite{assadi2025faster}), $\tilde O (mn^{1/3})$(Bhattacharya et al.~\cite{DBLP:conf/focs/BhattacharyaCCS24}), and $\tilde O (mn^{1/4})$ (Bhattacharya et al.~\cite{bhattacharya2025even}).
Duan et~al.~\cite{DBLP:conf/soda/DuanHZ19} further reduced the time to $O(m\cdot \poly(\log n, \epsilon^{-1}))$ as long as $\Delta= \Omega(\log^2n\cdot \epsilon^{-2})$, but use up to $(1+\epsilon)\Delta$ colors.
For bipartite graphs, Cole~\etal{}~\cite{bipartitecolouring} 
gave an optimal algorithm with $O(m \log \Delta)$ running time.
Cohen~\etal{}~\cite{DBLP:conf/focs/CohenPW19} recently studied the
problem in the online setting and proved various competitive
ratio results.

For dynamic graphs, Bhattacharya~\etal{}~\cite{deltaedgecolouring}
show how to maintain a $(2\Delta-1)$-edge coloring in $O(\log n)$ worst-case update time,
and that a $(2+\epsilon)\Delta$-edge coloring can be
maintained with $O(1/\epsilon)$ expected update time.
If $\Delta= \Omega(\log^2n\cdot \epsilon^{-2})$, Duan~\etal{}~\cite{DBLP:conf/soda/DuanHZ19} maintain an edge-coloring using $(1+\epsilon)\Delta$ colors in amortized $O(\log^8n\cdot \epsilon^{-4})$ update time.

\textbf{Maximum $k$-Edge Coloring.}
The problem was first studied
by Favrholdt and Nielsen~\cite{DBLP:journals/algorithmica/FavrholdtN03}
in the online setting, who
show that every algorithm that never chooses to not color (``reject'') a colorable edge
has a competitive ratio between $1/0.4641$ and $2$,
and that any online algorithm is at most
$\frac{4}{7}$-competitive.
Feige~\etal{}~\cite{originalpaper} showed %
that for every $k \geq 2$, there exists an $\epsilon_k > 0$ such that
it is NP-hard to approximate the problem within a ratio better than $(1+\epsilon_k)$.
They also describe a static $(1-(1-1/k)^k)^{-1}$-approximation algorithm for general $k$.
The currently best approximation ratios are $1/0.842$ for $k=2$ and $\frac{15}{13}$ for $k=3$~\cite{DBLP:journals/dam/ChenKM10,DBLP:journals/siamdm/KaminskiK14}.

The maximum $k$-edge coloring problem was first studied in the edge-weighted setting by Hanauer~\etal{}~\cite{infocomstatic}.
Here, instead of finding a maximum-cardinality subset of the edges, the total weight
of the colored edges is to be maximized.
In a follow-up work, Hanauer~\etal{}~\cite{infocomdynamic}
design a collection of different dynamic and batch-dynamic algorithms for weighted $k$-edge coloring.
Their focus is again more on the practical side.
Ferdous~\etal{}~\cite{DBLP:conf/esa/FerdousSPHK24} recently studied the problem in the streaming setting.

\textbf{Matching.}
The matching problem has been subject to extensive
research both in the unweighted and weighted case~\cite{DBLP:conf/focs/MicaliV80,DBLP:journals/jacm/GabowT91,DBLP:journals/siamcomp/HopcroftK73,DBLP:conf/soda/Gabow90,DBLP:conf/focs/DuanP10}.
Various dynamic algorithms with different trade-offs between update time and approximation ratio also exist for general~\cite{DBLP:conf/stoc/OnakR10,DBLP:conf/focs/BaswanaGS11,DBLP:conf/focs/Solomon16,DBLP:conf/stoc/NeimanS13,DBLP:conf/soda/BhattacharyaHI15,DBLP:conf/stoc/BhattacharyaHN16,oneplusepsilonmatching} and bipartite graphs~\cite{DBLP:conf/focs/BosekLSZ14,DBLP:conf/focs/ChenKLPGS22}.
Wajc~\cite{DBLP:conf/stoc/Wajc20} gives a metatheorem for rounding a dynamic
fractional matching against an adaptive adversary and a $(2 +
\epsilon)$-approximate algorithm with $O(1)$ update time or $O(\poly(\log n,
\epsilon^{-1}))$ worst-case update time.

\textbf{$\bvec{b}$-Matching.}
Gabow~\cite{gabowbmatch} gives a $O(\sqrt{\lVert b\rVert_1}m)$-time algorithm to compute a $\bvec{b}$-matching in the unweighted, static setting,
and an $O(\lVert b\rVert_1\cdot\min(m\log n, n^2))$-time algorithm for weighted graphs.
Ahn and Guha's algorithm~\cite{bmatchingapprox} computes a $(1+\epsilon)$-approximation for $\bvec{b}$-matching and runs in $O(m \poly(\log n, \epsilon^{-1}))$ time. Bienkowski et al.~\cite{DBLP:conf/sc/BienkowskiF023} give an online $O(\log b)$-approximate solution, which is asympotically optimal in the online setting.

For dynamic graphs, Bhattacharya~\etal{}~\cite{primaldual}
give a deterministic algorithm that maintains an $O(1)$-approximate fractional $k$-matching
with $O(\log^3 n)$ amortized update time.
This is improved by Bhattacharya~\etal{}~\cite{BhattacharyaGM17}, who show how to maintain an integral
$(2+\epsilon)$-approximate $\bvec{b}$-matching in expected amortized $O(1/\epsilon^4)$
update time against an oblivious adversary.

%% file: 11-Technical_Overview.tex
\section{Technical Overview}\label{sec:tech}
Our starting point is the following observation: The two problems $\bvec{b}$-matching and $k$-edge coloring seem very similar if $\bvec{b}$ is the vector having $b_v = k$ for every vertex $v \in V$.
Indeed, %
the colored edges in a $k$-edge coloring form a $k$-matching.
Vice versa, a maximum $k$-matching always contains a $\frac{k+1}{k}$-approximate $k$-edge coloring, as one can always color the edges with $k+1$ colors and discard the least-used color.
This close connection is one major ingredient for the two main algorithms we present, the dynamic \MatchO{} and \MatchA{} algorithms: Find a good $k$-matching in the graph first, then color it using as few colors as possible, and finally discard all edges of the least-used colors.
The goal is to perform updates in $O(\poly(\log n, \epsilon^{-1}))$ time.

Our \MatchO{} algorithm, which works against oblivious adversaries, is a combination of known algorithms:
We use the aforementioned $(2+\epsilon)$-approximation by Bhattarcharya et al.~\cite{BhattacharyaGM17} to find a $k$-matching. %
Duan et al.'s algorithm~\cite{DBLP:conf/soda/DuanHZ19} colors the edges with $(1+\epsilon)\Delta$ colors.
Discarding the least-used colors yields a $(2+\epsilon)\frac{k+1}{k}$-approximation for \MkEC{}.

For our $\MatchA{}$ algorithm, which is designed to work against an adaptive
adversary, we could have used the so-far best algorithm for $\bvec{b}$-matching by
Bhattacharya~\etal{}~\cite{primaldual}, which however only guarantees an
$O(1)$-approximation. %
The algorithm is based on a $(7+\epsilon)$-approximate \emph{fractional}
$\bvec{b}$-matching algorithm, whose solution is rounded.
We present an alternative rounding approach and thus obtain an improved integral
$k$-matching algorithm  that guarantees a
$(7+\epsilon)\frac{3k}{3k-1}$-approximation, which is at most $(8.4+\epsilon)$ (for $k=2$).
Following the same scheme as for \MatchO{}, we thus obtain a $(7+\epsilon)\frac{3k+3}{3k-1}$-approximation for \MkEC{}.

Our new rounding technique works as follows:
Given a fractional $k$-matching, we partition the edges of our graph by powers of $(1+\epsilon)$ according to their fractional value and maintain a $3\Delta$-coloring of each subgraph (which is not related to the coloring we will output).
We then construct a sparsified graph by choosing a subset of colors uniformly at random and keeping only edges of these colors.
Intuitively, the sparsified graph consists of those edges with high fractional values.
We show that if an optimal $k$-edge coloring of the input graph colors $s$ edges, then the sparsified graph contains at most $O(s \cdot\poly (\log n, \epsilon^{-1}))$ edges.
Running Ahn and Guha's algorithm~\cite{bmatchingapprox} on the sparsified graph takes $O(s \cdot\poly (\log n, \epsilon^{-1}))$ time and computes an integral $k$-matching.
We can thus afford to rerun Ahn and Guha's algorithm every $\Omega(\epsilon s)$ updates and still have polylogarithmic update time.
Recomputing the rounding this often also ensures the approximation ratio is still good enough.
We further prove that the integrality gap of the $k$-matching problem is $\frac{3k}{3k-1}$ and hence small.
This ensures that the sparsified graph, which we show contains a large fractional $k$-matching, also contains a large integral $k$-matching, and thus that the graph output by Ahn and Guha's algorithm is a good rounding of the original fractional matching.

We also show that the integrality gap of $\bvec{b}$-matching is small.
The argument starts by noting that the integrality gap of $\bvec{b}$-matching on {\em bipartite} graphs is $1$, which is a known result.
We then prove that every $\bvec{b}$-matching polytope is half-integral.
To do so, we build a bipartite graph that encodes the original graph.
Every extremal point of the polytope of the original graph can be encoded as half the sum of two extremal points of the bipartite polytope, and thus is half-integral.
Once we have a half-integral solution, we proceed to show that we can round it to an integer solution without losing much of the fractional solution.
Essentially, we look at the dual variables of the $\bvec{b}$-matching, that is, for each vertex, we look at its weighted degree.
We then show that while rounding, out of every carefully chosen three vertices, only one can see its degree drop, under the crucial condition that its weighted degree was already equal to $b_v$.
This allows us to prove that the integrality gap is $\frac{3\beta}{3\beta-1}$, with $\beta = \min_{v\in V} b_v$.

On bipartite graphs, many aspects of the problem become easier: the integrality gap of the $b$-matching algorithm, as well as the existence of efficient edge-coloring algorithms, like the one by Cole, Ost, and Schirra~\cite{bipartitecolouring}, which colors all of the edges with $\Delta$ colors in $O(m\polylog n)$ time.
This improves the approximation ratio of \MatchO{} and \MatchA{} to $(2+\epsilon)$ and $7(1+\epsilon)$ respectively.

%% file: 10-preliminaries.tex
\section{Preliminaries}\label{sect:prelim}
Unless denoted otherwise, we consider an undirected, unweighted graph $G = (V, E)$ with $n := |V|$
and $m := |E|$.
$G$ is dynamic, that is it undergoes an a priori unknown series of \emph{updates} in the form of \emph{edge insertions} and \emph{edge deletions}.
The \emph{update time} of an algorithm is the time it needs to process an update before accepting the next one.
These updates are controlled by an adversary, that can be either \emph{adaptive}, that is, can see the internal state of our data structure and choose the next update accordingly, or \emph{oblivious}, that is, has to decide all of the updates before we start running our algorithm.
$G$, $n$, and $m$ always refer to the current graph and its number of vertices and edges, respectively, i.e., including all updates that occurred beforehand.
Edges are treated as subsets of vertices of size $2$.
We use $\Delta(G)$ to denote the maximum vertex degree in $G$. If it is clear from the context, we may omit $G$ and just write $\Delta$.

\begin{definition}[Edge Coloring]
    Let $G=(V,E)$ be a graph, $k \in \N_+$, and $f: E\mapsto [k]\union \{\bot\}$ an edge coloring of the edges of $G$. We say that:
    \begin{enumerate}
        \item $f$ is a \emph{proper} coloring of $E$ if for any adjacent edges $e, e'$, we have that either $f(e) \ne f(e')$, or $f(e)=f(e')=\bot$.
        \item $f$ is a \emph{total} coloring of $E$ if $f^{-1}(\bot) = \varnothing$. We say it is \emph{partial} to emphasize it is not necessarily total, that is, $f^{-1}(\bot) $ may or may not be equal to $\varnothing$.
        \item A color $c \in [k]$ is \emph{free} on node $v$ (according to $f$) if no edge incident to $v$ has color $c$, that is, for every $e \ni v, f(e) \ne c$.
        \item An edge $e$ is \emph{colored} if $f(e) \in [k]$ and \emph{uncolored} otherwise.
    \end{enumerate}
\end{definition}

Given $G = (V, E)$ and $k \in \mathbb{N}$, the \emph{maximum $k$-edge coloring problem} consists in finding a proper coloring $f: E\mapsto [k]\union \{\bot\}$ of $E$ such that the set of colored edges $|f^{-1}([k])|$ is maximized.
In the rest of this paper, unless specified otherwise, every ($k$-)coloring is proper and we use $f$ to denote an arbitrary (proper) ($k$-)coloring, and $f^*$ to denote an optimal ($k$-)coloring.

A \emph{matching} $M \subseteq E$ is a subset of edges such that for every distinct pair $e, e' \in M$, $e \cap e' = \varnothing$.
Note that given a $k$-edge coloring $f$, $f^{-1}(c)$ is a matching for each $c \in [k]$.
A set of edges $M \subseteq E$ is a \emph{$k$-matching} if
for all $v \in V$, $|\{e \in M \colon e \ni v\}| \leq k$.
Given an $n$-dimensional vector $\bvec{b}^V$, we say that a set of edges $M \subseteq E$ is a \emph{$\bvec{b}$-matching} if for every $v \in V$, $|\{e \in M \colon e \ni v\}| \leq b_v$.
Thus, a $k$-matching is a $\bvec{b}$-matching where $\bvec{b} = k^V$, and a matching is a $1$-matching ($k = 1$).

\begin{theorem}[Coloring an Approximate $k$-Matching, extension of~\cite{originalpaper}]\label{thm:matchingcoloring}
    Let $G$ be a graph and $H$ a subgraph such that %
    $H$ is a solution of an $\alpha$-approximation algorithm for $k$-matching on $G$, %
    for some $\alpha \ge 1$. Let $f$ be a total coloring of $H$ using $k+\ell$ colors, with $\ell \in \N$. Then, discarding the $\ell$ least used colors from $f$ yields an $(\alpha\cdot \frac{k+\ell}{k})$-approximate $k$-edge coloring of $G$. In particular, if $k=\Delta(H)$ and $\ell = \epsilon \Delta(H)$, the approximation ratio is $\alpha\cdot( 1+\epsilon)$.
\end{theorem}

\begin{corollary}\label{lem:rel}
Given a graph $G$,
let $s^*$ be the size of an optimum $k$-matching in $G$ and let $p^*$ be the size of an optimum $k$-edge coloring. Then $p^*\le s^* \le \frac{k+1}{k} p^*$.
\end{corollary}

%% file: 5-Matching_polytope.tex
\section{\texorpdfstring{The $b$-Matching Polytope}{The b-Matching Polytope}}\label{sect:polytope}

In this section, we will show the following theorem.

\begin{theorem}[name=Integrality Gap Theorem,label=thm:integrality-gap,restate=integralitygaptheorem]
    The integrality gap of the $\bvec{b}$-matching polytope is $\frac{3\beta}{3\beta-1}$, where $\beta=\min_{v \in V} b_v$. The integrality gap of the bipartite $\bvec{b}$-matching polytope is 1.
\end{theorem}

While the result on bipartite graphs was known~\cite{schrijver2003combinatorial}, the general result does not exist in the literature to the best of our knowledge.
We first show %
that the (fractional) $\bvec{b}$-matching polytope is half-integral%
\footnote{Even though the $\bvec{b}$-matching polytope is well-researched, we could not find this result in the literature.}%
, then round an optimal solution for the fractional $\bvec{b}$-matching problem to get an integral solution with a good approximation ratio.

In this section, a \emph{trail} is a walk with no repeated edges (but possibly repeated nodes), a \emph{circuit} is a closed trail, and an \emph{Eulerian circuit} is a circuit that visits all edges.

The $\bvec{b}$-matching polytope is defined as follows:

\begin{definition}[fractional $\bvec{b}$-matching polytope]
Let $G=(V,E)$ be an undirected graph. The fractional \emph{$\bvec{b}$-matching polytope} $\Pea(G)$ is:
\[
\Pea(G)=\{\mathbf{x} \in {[0,1]}^E \colon \forall v \in V \sum_{e \ni v}x_e \le b_v\}
\]
\end{definition}

It is well-known that if $G$ is bipartite, then $\Pea(G)$ is integral, i.e., every vertex of the polytope has integer entries:
\begin{lemma}[\cite{schrijver2003combinatorial}]\label{lem:bipartiteintegral}
Let $G=(V,E)$ be a bipartite graph. Then $\Pea(G)$ is integral.
\end{lemma}

We will now show that the fractional $\bvec{b}$-matching polytope over general graphs is half-integral, that is, the entries of all its vertices are in $\{0, \frac{1}{2}, 1\}$. To do so, given a graph $G$, we build a graph $G'$ that is bipartite and show a relationship between vertices of $\Pea(G)$ and $\Pea(G')$.

\begin{lemma}\label{lem:halfintegral}
Let $G=(V,E)$ be a graph. Then $\Pea(G)$ is half-integral.
\end{lemma}

Next, we introduce a technique for rounding an optimal solution in $\Pea(G)$ to an integer solution that has similar total cost. 
This requires the following helper lemma:

\begin{lemma}\label{lem:noevencircuits}
    Let $G$ be a graph that contains no even circuit. Then $G$ has no even cycles, and no two odd cycles in $G$ share a node. %
\end{lemma}

We also recall the Euler partition of a graph:

\begin{definition}[Euler Partition]
Let $G=(V,E)$ be a graph. A \emph{Euler Partition} of $G$ is a partition of its edges into trails and circuits such that every node of odd degree is the endpoint of exactly one trail, and every node of even degree is the endpoint of no trail. 
\end{definition}

It is easy to see that such a partition exists for every graph, as one can compute one by removing maximal trails from $G$ until no edge remains. We now have all the tools necessary to find the integrality gap of the $\bvec{b}$-matching polytope, to which we give the proof sketch here, and the full proof in the appendix.

\integralitygaptheorem*

\begin{proof}[Proof Sketch]
Since the fractional polytope is half-integral, we find an optimal fractional solution $\mathbf{x}$ where all entries are in $\{0, \frac{1}{2}, 1\}$.
We then consider a Euler partition of the union of edges whose value is $\frac{1}{2}$. %
First, consider all trails that start and end at different vertices, and alternatively round up and down the values of the edges.
This does not affect the optimality of the solution: for each inner vertex of the trail, the same number of incident edges is rounded up and down. For each end vertex $x$, $\sum_{e \ni v}x_e \le b_v - \frac{1}{2}$, so rounding up does not violate the condition at $x$.

Note that all trails must have even length, otherwise $\mathbf{x}$ would not be optimal.
We again remove all integral edges from the subgraph, and end up with a Eulerian graph.
We then apply the same strategy to every even circuit in the subgraph, until there are no more even circuits. 
By Lemma~\ref{lem:noevencircuits}, the remaining graph is composed of disjoint odd cycles. 
We can thus again apply the strategy of alternatively rounding up and down the values of the edges, except that this time, there might be two consecutive edges we might need to round down, %
which may decrease the solution value.
The worst-case scenario occurs for length-$3$ cycles, as the rounded solution is only $2/3$ as good as the fractional solution.
But this only happens when no node on the cycle can ``afford'' both incident edges to be rounded up, that is, all dual inequalities are tight to a $1/2$ additive factor.
This gives the result.
\end{proof}

%% file: 7-sparsificationscheme.tex
\section{The Sparsification Scheme}\label{sec:sparsification}
Given a dynamic fractional $k$-matching algorithm $\Alg_m$, this section provides a sparsification scheme for fractional $k$-matching that enables us to give an algorithm $\Alg$ that maintains an approximate integral $k$-matching:

\begin{theorem}[Sparsify and Round]\label{thm:rounded}
    Let $G$ be a dynamic graph, $\epsilon>0$, $\alpha=O(1)$, and $\Alg_m$ a dynamic algorithm that maintains an $\alpha$-approximate fractional $k$-matching of $G$ in $O(T_m)$ update time.
Then there exists an algorithm $\Alg$ that maintains an $\alpha (1+\epsilon) \frac{3k}{3k-1}$-approximate integral $k$-matching in $O(T_m + \poly(\log n, \epsilon^{-1}))$ amortized update time against an adaptive adversary.
 If $G$ is bipartite, the approximation ratio reduces to $\alpha (1+\epsilon)$.
\end{theorem}

This result will be particularly useful for the \MatchA{} algorithm in Section~\ref{sect:matcha}.
To prove the theorem, we consider a dynamic graph $G$ and assume we have access to a dynamic  algorithm $\Alg_m$ that has update time $O(T_m)$ and maintains a fractional $k$-matching $\mathbf{x}$ with approximation factor $\alpha$. The goal is to maintain a sparse subgraph $H$ of $G$ that contains an integral $k$-matching whose size is within an $\alpha(1+\epsilon)\frac{3k}{3k-1}$ factor of the size of an optimal $k$-matching in $G$. 

To this end, we separate our strategy into two schemes: the update scheme is repeated at every update, while the request scheme is only repeated when we need the sparse graph $H$.
Our technique is an adaptation and generalization of an efficient sparsification by Wajc~\cite{DBLP:conf/stoc/Wajc20} for rounding fractional (1-)matchings to integral ones. 

\paragraph*{Update Scheme} Let $\mathbf{x}$ be the fractional $k$-matching maintained by $\Alg_m$ of value $c(\mathbf{x})=\sum_{e \in E} x_e$.
Let $\ell := 2\log_{(1+\epsilon)}(n \epsilon^{-1})$.
We maintain $\ell$ subgraphs $G_1, \dots, G_\ell$ of $G$, where $G_i = (V, E_i)$ and
\[
E_i= \left\{e \in E \colon x_e \in \big((1+\epsilon)^{-i}, (1+\epsilon)^{-i+1}\big]\right\}
\]

Let $E^+=\Union_{i \in [\ell]}E_i$.
Note that $E^+$ is a subset of $E$ and does not contain edges $e$ such that
$x_e \leq (1+\epsilon)^{-\ell} = \frac{\epsilon^2}{n^2}$.
Hence, 
$\sum_{e \in E \setminus E^+} x_e \leq m \frac{\epsilon^2}{n^2} \leq \epsilon^2$ and we
obtain
\begin{equation}\label{eq:sumeplus}
\sum_{e\in E^+}x_e \ge c(\mathbf{x}) - \epsilon^2 \ge c(\mathbf{x})(1-\epsilon).
\end{equation}

In each $G_i$, we maintain a proper, total $(3\ceil{k(1+\epsilon)^i})$-edge coloring. Since in $G_i$, every edge satisfies 
$x_e > (1+\epsilon)^{-i}$,
and $\mathbf{x}$ is a $k$-matching, we know that the maximum degree of $G_i$ can be no higher than $k(1+\epsilon)^i$. Hence, every edge modified in $G_i$ can be colored in expected constant time\footnote{By maintaining a hash table at each vertex of the free colors, since we have more than $3\Delta (G_i)$ colors available. By picking a color at random, we have a probability higher than $\frac 1 3$ for this color to be free at both end nodes, and thus we only need three random picks in expectation to find a free color.}.
As each update of $G$ can modify only $O(T_m)$ edges, there are at most $O(T_m)$ modifications in total to all subgraphs $G_i$.
The recoloring caused by each modification can be handled in expected constant time, which implies $O(T_m)$ expected time per update to $G$.

\paragraph*{Request Scheme} We fix a parameter $d:=\max\{\frac 1 {k\epsilon}, \frac{4\log(2/\epsilon)}{k\epsilon^2}\}$.
Whenever we need access to the sparse graph $H$, we obtain a set of edges $H_i$ from each $G_i$ by choosing uniformly at random without replacement up to $\ceil{kd(1+\epsilon)}$ colors, and setting $H_i$ to be the set of edges colored with those colors.
Then, $H := (V, \Union_{i=1}^\ell H_i)$.
Obtaining $H$ takes $O(\card H)$ time. 

We analyze the size of $H$ with respect to $\card{{f^*}^{-1}([k])}$, the size of an optimal $k$-edge coloring in $G$. Let $p^* := \card{f^{*-1}([k])}$ and $s^*$ be the size of a maximum $k$-matching on $G$. Each collection of $k$ colors in $H_i$ creates a $k$-edge coloring in $G$, and has thus size at most $p^*$. We have $O(d(1+\epsilon))$ such collections in $H_i$, and hence $\card{H_i}=O(d(1+\epsilon)p^*)=O(d(1+\epsilon)s^*)$. Therefore,
\[
\card{H} \le \sum_{i =1}^{\ell} \card{H_i} \le O\left(\ell d(1+\epsilon)s^*\right)%
=O\left(s^*\cdot \log n\poly(\epsilon^{-1})\right).
\]

Together with Equation~(\ref{eq:sumeplus}), Lemmas~\ref{lem:sparsification1} and~\ref{lem:sparsification} below show that in expectation, $H$ contains
a fractional $k$-matching of total value at least $c(\textbf{x})(1-\epsilon)(1-6\epsilon)\ge c(\textbf{x})(1-7\epsilon)$. By our Integrality Gap Theorem (Theorem~\ref{thm:integrality-gap}), $H$ then contains
an integral $k$-matching of cardinality greater than $c(\textbf{x})(1-7\epsilon)\frac{3k-1}{3k}$. Since the fractional dynamic algorithm outputs an $\alpha$-approximation $c(\textbf{x})$ of the optimal fractional solution, we have that $s^* \le \alpha c(\textbf{x})$.
Therefore $H$ contains a $k$-matching of cardinality greater than $\frac {3k-1} {3k} (1-7\epsilon)s^* / \alpha$. We thus get the following theorem:

\begin{theorem}[Sparsification]\label{thm:sparsification}
    Let $G$ be a dynamic graph, $\epsilon>0$, and $\Alg_m$ a dynamic algorithm that maintains an $\alpha$-approximate fractional $k$-matching of $G$ in $O(T_m)$ update time.
    Let $s^*$ be the size of an optimal $k$-matching in G.
    Then the sparsification scheme maintains a sparsification $H$ of $G$ that runs in $O(T_m)$ update time and $O(s^* \cdot \poly(\log n, \epsilon^{-1}))$ request time. 
    In expectation, $H$~contains an integral $k$-matching
    of size at least $s^*/\left(\alpha \frac{3k}{3k-1} (1+\epsilon)\right)$ and satisfies $\card H =O(s^* \log n \poly(\epsilon^{-1})) $. If $G$ is bipartite, the approximation ratio reduces to $\alpha (1+\epsilon)$.
    \end{theorem}

    We now prove the Sparsify and Round Theorem: 
    
    \begin{proof}[Proof of Theorem~\ref{thm:rounded}]
        We build the
    algorithm $\Alg$ that maintains an approximate integral $k$-matching as follows: We use $\Alg_m$ to maintain an $\alpha$-approximation of a fractional $k$-matching in $G$. At a given update, we compute a sparsification $H$ of $G$ using Theorem~\ref{thm:sparsification} 
of size $O(s^*\poly(\log n, \epsilon^{-1}))$ that contains an integral $k$-matching of size at least 
$\frac{1}{\alpha \frac{3k}{3k-1} (1+\epsilon)}\cdot s^*$.
We then run the Ahn-Guha algorithm~\cite{bmatchingapprox} on $H$ in $O(s^* \poly(\log n, \epsilon^{-1}))$ time to get a 
$\alpha \frac{3k}{3k-1} (1+O(\epsilon))$-approximate $k$-matching
$s$ of $G$. Since every update only changes the size of an optimal solution by at most $1$, the computed $k$-matching remains a good approximation of the optimal solution over the next 
$\epsilon s$ updates, which yields an amortized update time of $O(\poly(\log n, \epsilon^{-1}))$. 
\end{proof}

Next, we state Lemmas~\ref{lem:sparsification1} and~\ref{lem:sparsification}, whose proofs are given in the appendix.

\begin{lemma}\label{lem:sparsification1}
Let $i \in \N$, and $G_i=(V,E_i)$ be a graph. Let $\textbf{x}$ be a fractional $k$-matching on $G_i$ that satisfies $x_e \in \big((1+\epsilon)^{-i}, (1+\epsilon)^{-i+1}\big]$, and $d\ge \frac 1 {k\epsilon}$.
If $d\ge (1+\epsilon)^{i-1}$, set $H_i := E_i$.
Otherwise, let $f_i$ be a total $(3\ceil{k(1+\epsilon)^i})$-edge coloring of $G_i$. 
 Sample $3\ceil{kd}$ colors uniformly at random (without replacement), and set $H_i$ to be the set of all edges colored by one of the sampled colors.
 Then each edge $e$ is sampled with probability $\Proba[e \in H_i]$ such that
\begin{equation}
\min\{1, x_e \cdot d\}\cdot (1+\epsilon)^{-2}\le \Proba[e \in H_i]\le \min\{1, x_e \cdot d\}\cdot (1+\epsilon). \label{eq:probainH}
\end{equation}

Then, if $x_e >\frac 1 d$, $\Proba[e \in H_i]=1$. Moreover, any two adjacent edges are negatively associated, that is, for any edges $e$ and $e'$ that share a node, we have $\Proba[X_e|X_e']\le \Proba[X_e]$.
Furthermore, the random variables $\{[X_e|X_{e'}] \mid e \ni v\}$ are negatively associated for any $v\in V, e' \ni v$. 
\end{lemma}

%% file: 8-sparsificationlemma.tex
\begin{lemma}[Sparsification]\label{lem:sparsification}
Let $\textbf{x}$ be a fractional $k$-matching on a graph $G$, $\epsilon \in (0, \frac 1 2)$, and $d\ge \max \left\{\frac 1 {k\epsilon}, \frac{4\log(2/\epsilon)}{k\epsilon^2}\right\}$. Let $H$ be a subgraph of $G$, where each edge of $G$ is sampled with probability $\Proba[ e \in H]$, where $\Proba[e \in H]=1$ if $x_e > \frac 1 d$ and
\begin{equation}\label{eq:probestimate}
    \min\{1, x_e \cdot d\} \cdot (1+\epsilon)^{-2} \le \Proba[e \in H] \le \min\{1,x_e \cdot d\}\cdot (1+\epsilon).
\end{equation}
Let $X_e:= \1[e \in H]$. The edges need not be sampled independently, however two edges that are adjacent need to be negatively associated, that is, for any edges $e$ and $e'$ that share a node, we have $\Proba[X_e | X_e']\le \Proba[X_e]$. We also require that for any $v\in G, e' \ni v$, the random variables $\{[X_e|X_{e'}] |e \ni v\}$ are negatively associated. 
Then, $H$ has a fractional $k$-matching $\mathbf{y}$ of expected value at least
   \[ 
    \expect\left[\sum_e y_e\right]\ge \sum_e x_e(1-6\epsilon).
   \] 
\end{lemma}

%% file: 4-Dynamic.tex
\section{\texorpdfstring{Dynamic Algorithms for Maximum $k$-Edge Coloring}{Dynamic Algorithms for Maximum k-Edge Coloring}}\label{sect:dynamic}

\subsection{The \Greedy{} Algorithm}\label{sect:greedy-dyn}
Our first algorithm follows a simple greedy scheme: %
If an edge is added to the graph, we only check whether there is a common free color available at both its end nodes, and if it is the case, we color this edge with that color, otherwise we do nothing. 
If an edge colored $c$ is removed from the graph, we try to color one edge adjacent to each of its end nodes with color $c$. This keeps the coloring maximal, and is thus a $(1+2\frac {\sqrt{3}}3)$-approximation, as shown by Favrholdt and Nielsen~\cite{DBLP:journals/algorithmica/FavrholdtN03}.
If we maintain a table of free colors at every vertex, each insertion takes $O(\min\{k, \Delta\})$ time, as we have to go through all used colors at the endpoints to find a free one.
Every edge deletion takes $O(\Delta)$ time, as we have to check whether the color $c$ is free at the end of $2\Delta$ edges in the worst case.

\begin{theorem}[Greedy Algorithm]\label{thm:greedy}
     The \Greedy{} algorithm is deterministic and maintains a $(1+2\frac {\sqrt{3}}3\approx 2.155)$-approximation of a maximum $k$-edge coloring in $O(\Delta)$ update time. %
\end{theorem}

Note that in the case of bounded arboricity, one can maintain an orientation of the graph such that the out-degree of each node is at most a multiple of the arboricity, as shown by Chekuri~\etal{}~\cite{DBLP:conf/soda/ChekuriCHHQRS24}, in polylogarithmic time.
In this case, each vertex is ``responsible'' for letting neighboring out-vertices know about its available colors.
In that case, the update time drops to $O(c)$, where $c$ is the arboricity of the graph, as the limiting factor now is not finding a suitable color for an edge, but rather updating the available colors on each vertex.

\subsection{The Amortized Algorithms}

Our next two algorithms rely on $k$-matchings. The idea is to maintain an approximate $k$-matching, and have it totally colored. However, as coloring can be expensive, we will not use the coloring algorithm at every update, but rather only once over multiple rounds, so we can amortize its cost. 
The following lemma formalizes the amortization technique:

\begin{lemma}[Amortization]\label{lem:amortize}
Let $\epsilon>0$, and assume we compute a coloring $f$ that colors $p:=\card{f^{-1}([k])}$ edges of a graph $G$. 
Then assume we have up to $\floor{\epsilon p}$ edge insertions and deletions to $G$.
Define a coloring $g$ on $G$ as follows: $g(e):=f(e)$ if $e$ was already in $G$ before the modifications, and $g(e):=\bot$ otherwise. Then if $f$ is an $\alpha$-approximation of maximum $k$-edge coloring before the updates, $g$ is a $\alpha(1+3\epsilon)$-approximation after the updates if $\epsilon \le \frac 1 3$.
\end{lemma}

\subsubsection{The \MatchO{} Algorithm}\label{sect:matcho}

If the updates to the graph are controlled by an oblivious adversary, we can use Bhattacharya, Gupta, and Mohan's algorithm~\cite{BhattacharyaGM17} for dynamic $\bvec{b}$-matching. They maintain an integral $(2+\epsilon)$-approximation of $\bvec{b}$-matching against an oblivious adversary in $O(\epsilon^{-4})$ update time. 

Our \MatchO{} algorithm works as follows: We use Bhattacharya~\etal{}'s algorithm to maintain a $(2+ \epsilon)$-approximation for $k$-matching, which is represented by a graph $H$. Then, we compute a $(k+1)$-edge coloring using Gabow's coloring algorithm~\cite{gabowcolor}
on $H$, and discard the least used color to obtain a $(2+ \epsilon )\frac{k+1}{k}$-approximation $f$ of a $k$-edge coloring, guaranteed by Theorem~\ref{thm:matchingcoloring}.
We refrain from recomputing the coloring for the next $\floor{ \epsilon  \card{f^{-1}([k])}}$ updates, while continuing to update the $k$-matching after each update operation. By the Amortization Lemma (Lemma~\ref{lem:amortize}), this yields a $(1+ 3  \epsilon)(2+\epsilon )\frac{k+1}{k}$-approximation, which is a $(2+5\epsilon)\frac{k+1} k$-approximation if $\epsilon \le \frac 1 3$. This is particularily efficient if $\Delta(H) = O(\log^2n \cdot \epsilon ^{-2})$.

If on the other hand $\Delta(H) = \Omega(\log^2n \cdot \epsilon ^{-2})$, using the edge-coloring by Duan, He, and Zhang~\cite{DBLP:conf/soda/DuanHZ19}, we color all the edges of $H$ with $(1+\epsilon)\Delta(H)$ colors. Discarding the $\epsilon\Delta(H)$ colors that color the fewest edges among them, this yields a $(2+O(\epsilon))$ approximation of the maximum $k$-edge coloring by Theorem~\ref{thm:matchingcoloring}.

\emph{Running time analysis.}
Let $s$ be the size of $H$ at the time of recoloring.
Computing the recoloring (whether it is a $(k+1)$ or $(1+\epsilon)\Delta(H)$-edge coloring) and removing the least used color(s) gives a $k$-edge coloring $f$ of size at least $\frac{k}{k+1}  s$.
Thus,
$s \le \frac {k+1}{k} \card{f^{-1}([k])} = O(\card{f^{-1}([k])})$.

Let us now analyze the amortized update time. For each update, we spend $O(\epsilon^{-4})$ time to maintain the $k$-matching. 
If $\Delta(H) = O(\log^2n \cdot \epsilon ^{-2})$, coloring the $k$-matching with Gabow's algorithm takes $O(s\cdot \Delta(H)\log n) = O(s\cdot\poly(\log n, \epsilon^{-1}))$ time. If on the other hand $\Delta(H) = \Omega(\log^2n \cdot \epsilon^{-2})$,
coloring the $k$-matching with Duan, He, and Zhang's algorithm takes $O(s\poly(\log n, \epsilon^{-1})) $ time. Either way, this can be amortized over the next $\floor{ \epsilon  \card{f^{-1}([k])}}$ updates, yielding an amortized update time of $O(\poly(\log n, \epsilon^{-1}))$ for the complete algorithm.

We hence have the following result:

\begin{theorem}[MatchO Algorithm]\label{thm:matcho}
    Against an oblivious adversary, the \MatchO{} algorithm runs in $O(\poly(\log n, \epsilon^{-1}))$ amortized update time and maintains a $k$-edge coloring that is in expectation a $(2+\epsilon)\frac{k+1} k$-approximation.
\end{theorem}

In the case of a bipartite graph, we can color the $k$-matching with $k$ colors in $O(m \log k)$ time using
the algorithm by Cole, Ost, and Schirra~\cite{bipartitecolouring} instead of either Gabow's or Duan, He, and Zhang's algorithm. This yields a better approximation ratio:

\begin{theorem}[Bipartite MatchO Algorithm]\label{thm:matcho-bip}
    Against an oblivious adversary, the \bMatchO{} algorithm runs in $O(\poly(\log n, \epsilon^{-1}))$ amortized update time and maintains a $k$-edge coloring that is in expectation a $(2+\epsilon)$-approximation.
\end{theorem}

\subsubsection{The \MatchA{} Algorithm}\label{sect:matcha}
If the updates to the graph are controlled by an adaptive adversary, we can use Bhattacharya, Henzinger and Italiano's~\cite{primaldual} algorithm for fractional $\bvec{b}$-matching. They maintain an $(7+\epsilon)$-approximate fractional $\bvec{b}$-matching of the dynamic graph in $O(\log (m+n)\epsilon^{-2})$ time against an adaptive adversary. 

Our \MatchA{} algorithm works as follows: We use Bhattacharya~\etal{}'s algorithm as described in Section~\ref{sec:sparsification} to maintain a sparsification $H$ of $G$. By the Sparsify and Round Theorem (Theorem~\ref{thm:rounded}), this takes $O(\log (m+n)\epsilon^{-2})$ update time and $O(\card{H})$  time to output $H$ when requested. 
Let $f^*$ be an optimal $k$-edge coloring of $G$ of size $p^* := \card{f^{*-1}([k])}$.
By Theorem~\ref{thm:rounded}, $H$ in expectation contains an integral $k$-matching of size at least $p^*/\left(7 \frac{3k}{3k-1} (1+\epsilon)\right)$ and $\card H =O(p^* \log n\cdot\poly(\epsilon^{-1}))$.

Similarly to the case with an oblivious adversary, we will only compute a coloring in few, carefully selected rounds, and thus will not need  to access $H$ in every round.
More specifically, we use the amortization technique from Lemma~\ref{lem:amortize} to determine in which rounds to recolor $H$ for the current graph. To recolor we run Ahn and  Guha's static algorithm for $\bvec{b}$-matching~\cite{bmatchingapprox} on $H$ to compute a $(1+\epsilon)$-approximate (integral) $k$-matching $H'$ of $H$.
This ensures that the sparse graph has degree at most $k$.
Finally, we either compute a $(k+1)$-edge coloring using Gabow's algorithm for edge-coloring~\cite{gabowcolor} of $H'$, if $\Delta(H') = O(\log^2n \cdot \epsilon ^{-2})$, or a $(1+\epsilon)\Delta(H)$-edge coloring using Duan, He, and Zhang's Algorithm otherwise, and discard the least used colors to obtain
a $7  (1+\epsilon)
\frac{3k}{3k-1}\frac{k+1}{k}$-approximation $f$ of the maximum cardinality $k$-edge coloring of $G$.

We refrain from recomputing the coloring for the next $\floor{ \epsilon  \card{f^{-1}([k])}}$ updates, while continuing to maintain the $k$-matching. By Lemma~\ref{lem:amortize}, this yields an $7(1+ 3  \epsilon) (1+\epsilon)\frac{3k+3}{3k-1}$-approximation $f'$ of the current $f^*$, which is an $7(1+ 5  \epsilon)\frac{3k+3}{3k-1}$-approximation if $\epsilon \le \frac 1 3$.

\emph{Running time analysis.}
Let us analyze the amortized update time. For each update, we must first spend $O(\log n\cdot \text{poly}(\epsilon^{-1}))$ time to maintain the fractional $k$-matching and its sparsification.  
Requesting the sparse graph $H$ takes $O(\card{H})$ time. Finding a $k$-matching in it takes $O(\card H \cdot \poly(\log n, \epsilon^{-1}))$ time~\cite{bmatchingapprox}. Let $s$ be the size of that $k$-matching after the current update. Coloring the $k$-matching with either Gabow's or Duan, He, and Zhang's algorithm takes $O(s\poly(\log n, \epsilon^{-1}))$ time. 
Thus the total time for all updates in an interval starting at a recoloring and containing all updates up to the next recoloring is $O((s+\card H) \cdot\text{poly} (\log n, \epsilon^{-1}))$.

Let $s^*$ be the size of the optimum $k$-matching after the current update. Recall that
$s^* \le \frac {k+1}{k} p^*$ by Corollary~\ref{lem:rel}.
Thus, we have that $s \le s^* \le \frac {k+1}{k} p^* = O(p^*)$.
We also have $\card H = O(p^* \log n\cdot\poly(\epsilon^{-1}))$.
Hence, the total update time in an interval is 
$O(p^* \cdot \text{poly} (\log n, \epsilon^{-1}))$

Note also that $\card{{f'}^{-1}([k])}=O(\card{f^{-1}([k])})$
as we recolor every $\floor{ \epsilon  \card{f^{-1}([k])}}$ updates and each update changes the size of the $k$-edge coloring by at most one. Thus it follows that
\[
p^* \le 7(1+ 5\epsilon)\frac{3k+3}{3k-1} \card{f'^{-1}([k])}=O(\card{f'^{-1}([k])})
=O(\card{f^{-1}([k])}).
\]
Hence we can amortize the total update time of an interval over the length of an interval, which consists of 
$\floor{ \epsilon  \card{f^{-1}([k])}}$ updates,
to achieve an amortized update time of $O(k \cdot \poly(\log n, \epsilon^{-1}))$, resulting in the following theorem.

\begin{theorem}[MatchA Algorithm]\label{thm:matcha}
    Against an adaptive adversary, the \MatchA{} algorithm runs in expected $O(\poly(\log n, \epsilon^{-1}))$ amortized update time and maintains a $k$-edge coloring that is in expectation a $7(1+ \epsilon)\frac{3k+3}{3k-1}$-approximation.
\end{theorem}

Similarly to the previous section, in the case of a bipartite graph, we can color the $k$-matching with $k$ colors in $O(m \log k)$ time using Cole, Ost, and Schirra's algorithm~\cite{bipartitecolouring} instead of either Gabow's or that of Duan, He, and Zhang. This also drops the $\frac{k+1} k$ factor in the approximation computation above. Moreover, the integrality gap for the $k$-matching becomes 1, removing the $\frac {3k}{3k-1}$ factor. This yields a better approximation ratio:

\begin{theorem}[Bipartite MatchA Algorithm]\label{thm:matcha-bip}
    Against an adaptive adversary, the \bMatchA{} algorithm runs in expected $O(\poly(\log n, \epsilon^{-1}))$ amortized update time and maintains a $k$-edge coloring that is in expectation an $7(1+ \epsilon)$-approximation.
\end{theorem}

%% file: 9-conclusion.tex
\section{Conclusion}\label{sect:conclusion}

In this work, we have initiated the study of fully dynamic approximation algorithms for the NP-hard $k$-edge coloring problem by presenting and analyzing three dynamic algorithms.
Moreover, we have demonstrated the close relationship between $\bvec{b}$-matching and $k$-edge coloring, making any advances in $\bvec{b}$-matching to automatically translate into better results for $k$-edge coloring. In the future, it would be thus interesting to investigate more into $\bvec{b}$-matching algorithms.
In particular there is space for improvement in finding dynamic (fractional or not) $\bvec{b}$-matching algorithms against an adaptive adversary with approximation ratio better than $7$ which still run in polylogarithmic time.

%% file: appendix.tex
\section{Omitted Proofs}\label{sect:app:proof:lem:amortize}

\begin{proof}[Proof of Theorem~\ref{thm:matchingcoloring}]
    Let $s^*$ be the size of an optimum $k$-matching in $G$ and let $p^*$ be the size of an optimum $k$-edge coloring. Since any $k$-edge coloring is a $k$-matching, we have that $p^*\le s^*$.
    Let $s=\card{H}$. The $\ell$ colors that color the smallest number of edges color at most $\frac \ell {k+\ell} s$ edges, otherwise the average of colored edges by color exceeds $\frac s {\ell+k}$. Let $p$ be the number of edges colored by the remaining colors. Then $p \ge \frac k {k+\ell} s \ge \frac k {k+\ell}\frac 1 \alpha s^* \ge \frac k {k+\ell}\frac 1 \alpha p^* $.
\end{proof}

\begin{proof}[Proof of Theorem~\ref{thm:integrality-gap}]
    Since the $\bvec{b}$-matching polytope is half-integral, we can find an optimal fractional $b$-matching $\mathbf{x}$ of $G = (V, E)$ such that $\mathbf{x}$ is half-integral.
    Let $H = (V, E_p)$ be the subgraph of $G$ with $E_p=\{e\in E \colon x_e \notin\{0,1\}\}$. If $E_p = \varnothing$, $\textbf{x}$ is integral.
    
    Otherwise, consider an Euler partition of $H$.
    If it contains a trail $T$ which starts and ends at different nodes, write $T=\{e_1, \dots, e_{\card{T}}\}$ where each $e_i$ is adjacent to $e_{i+1}$.
    Let $\mathbf{x^+}$ be defined by
\[
x^+_e :=\left\{\begin{array}{ccc}
    x_e & \text{if} & e \notin T, \\
    x_e + \frac 1 2(-1)^{i+1} & \text{if} & e=e_i,
\end{array}\right.
\]
i.e., we alternatingly round up and down by $\frac{1}{2}$ along the trail.
We clearly have that $\mathbf{x}^+ \in \Pea(G)$, since $\mathbf{x}\in [0,1]^m$ and for each node $v\in V$ except the endpoints of $T$, an equal number of edges incident to $v$
is rounded up and down.
An end point $v$ of the trail might have at most one more edge that sees its value increase than decrease by $\frac 1 2$. Since the node $v$ is of odd degree in $E_p$, we have that $\sum_{e \ni v} x_e \le b_v - \frac 1 2$, which implies $\sum_{e \ni v} x_e^+ \le b_v$, i.e. the condition is still satisfied. We moreover have that $\sum_e x_e \le \sum_e x^+_e$.

We can, hence, reduce the number of nodes of odd degree in $E_p$ by creating solutions that are as good as $\mathbf{x}$ that have fewer and fewer odd degree nodes. We end up with $H$ having only even degree nodes. By Euler's Theorem, there exists an Euler Circuit on every connected component of $H$.

For every node $u$, let $x(u):=\sum_{e\ni u}x_e$, and let $C$ be a connected component of $H$. If there exists a node $u \in C$ such that $x(u)<\bvec{b}_u$, then $x(u) \le \bvec{b}_u-1$, and we can write the connected component $C=\{e_1, \dots, e_{\card{C}}\}$ as an Eulerian circuit where each $e_i$ is adjacent to $e_{(i+1) \bmod \card{C}}$, and where $e_1$ is adjacent to $u$. Then define:
\[
x^+_e:=\left\{\begin{array}{ccc}
    x_e & \text{if} & e \notin C \\
    x_e + \frac 1 2(-1)^{i+1} & \text{if} & e=e_i 
\end{array}\right.
\]
We clearly have that $\mathbf{x}^+ \in \Pea(G)$, since $\mathbf{x}\in [0,1]^m$, and for each node $v\in V$ except for $u$, we have that the number of edges adjacent to $v$ losing $\frac 1 2$ is equal to the number of edges adjacent to $v$ winning $\frac 1 2$ canceling out. Node $u$ has at most two more edge that sees its value increase than decrease by $\frac 1 2$. Since $x(u) \le b_u - 1$, we have $\sum_{e \ni u} x_e^+ \le b_u $, the condition is still satisfied. We moreover have that $\sum_ex_e \le \sum_ex^+_e$.

We can also show that even if no $u \in C$ satisfies $x(u)<b_u$, but $C$ has an even number of nodes, the number of edges in the component is even, and thus the Eulerian circuit of $C$ is of even length, and thus computing $\mathbf{x}^+$ as above will yield an integer solution on $C$. Note as well that if $C$ is of size $1$, then $C$ contains no edges and $x$ is already integral on $C$.

We end up with connected components $C_1, \dots, C_\ell$ of odd size at least $3$, where every node $u \in \Union_{i \in [\ell]} C_i$ satisfies $x(u)=b_u$. Each $C_i$ can be seen as an Eulerian circuit. We will now build an integer solution $\mathbf{x}^-$ that approximates $\mathbf{x}$. For every $i \in [\ell]$, pick  an arbitrary node $u_i\in C_i$, and write $C_i=\{e_1^i, \dots, e^i_{\card {C_i}}\}$ where each $e^i_j$ is adjacent to $e^i_{j+1 \bmod \card{C}}$, and where $e^i_1$ is adjacent to $u^i$.
Define then:
\[
x^-_e :=\left\{\begin{array}{ccc}
    x_e & \text{if} & e \notin \Union_{i\in [\ell]}C_i \\
    x_e + \frac 1 2 (-1)^j & \text{if} & e=e^i_j 
\end{array}\right.
\]

We clearly have that $\mathbf{x}^- \in \Pea(G)$, since $\mathbf{x}\in [0,1]^m$ and for each node $v\in V$, we have that the number of edges adjacent to $v$ losing $\frac 1 2$ is larger than the number of edges adjacent to $v$ winning $\frac 1 2$, and thus $\sum_{e \ni v} x^-_e \le \sum_{e \ni v} x_e \le b_v$. We moreover have that
\begin{align*}
    \sum_ex^-_e &= \frac 1 2 \sum_{v\in V} \bvec{x}^-(v) = \frac 1 2 \sum_{v\in V\setminus \Union_{i\in [\ell]}C_i} x(v)+ \frac 1 2 \sum_{i \in [\ell]}\Big( x(u_i)-1+\sum_{v \in C_i, v\neq u_i} x(v)\Big)\\
    &\ge \frac 1 2 \sum_{v\in V\setminus \Union_{i\in [\ell]}C_i} x(v)+\frac 1 2 \sum_{i \in [\ell]}\Big(\bvec{b}_{u_i}-1+\sum_{v \in C_i, v\neq u_i} b_v\Big)
    \\&= \frac 1 2 \sum_{v\in V\setminus \Union_{i\in [\ell]}C_i} x(v)+\frac 1 2 \sum_{i \in [\ell]}\Big(-1+\sum_{v \in C_i} b_v\Big). 
\end{align*}

However, we have that $\sum_{v \in C_i} b_v \ge \card{C_i}\beta$ and that for each $i \in [\ell]$, $\card{C_i} \ge 3$ and thus $\ell \le \frac 1 3\card{\Union_{i \in [\ell]}C_i}$. Therefore:
\[
\frac 1 2 \sum_{i \in [\ell]}\Big(-1+\sum_{v \in C_i} b_v\Big)\ge -l+\frac 1 2 \sum_{i \in [l] }\sum_{v \in C_i}b_v\ge \big( 1-\frac 1 {3\beta}\big) \frac 1 2 \sum_{i \in [l]} \sum_{v \in C_i} b_v = \big( 1-\frac 1 {3\beta}\big) \frac 1 2 \sum_{i \in [l]} \sum_{v \in C_i} x(v)
\]
Hence,
\[
\sum_e x^-_e \ge \frac 1 2 \sum_{v\in V\setminus \Union_{i\in [\ell]}C_i} x(v) + \big( 1-\frac 1 {3\beta}\big) \frac 1 2 \sum_{i \in [l]} \sum_{v \in C_i} x(v) \ge \big( 1-\frac 1 {3\beta}\big) \frac 1 2 \sum_{v \in V} x(v).
\]
Since $\mathbf{x}^-$ is integral, the theorem follows.
\end{proof}

\begin{proof}[Proof of Lemma~\ref{lem:halfintegral}]
    Let us write $V=\{v_1, \dots, v_n\}$ and $E=\{e_1, \dots, e_m\}$. Define $G'=(V', E')$, where $V'=\{v'_1, \dots v_n', v''_1, \dots, v''_n\}$, and $E'=\{e_1', \dots, e_m', e''_1, \dots, e''_m\}$, where for each $i \in [m]$, if $e_i=(v_j, v_\ell)$ for some $j,\ell \in [n]$, then $e'_i$ and $e''_i$ are defined as $e'_i=(v'_j, v''_\ell)$ and $e''_i=(v''_j, v'_\ell)$. Define $b'$ such that $b'_{v_i'}=b'_{v_i''}=b_{v_i}$ for all $i$.

    Let $\mathbf{x}$ be a vertex of $\Pea(G)$. Then $\mathbf{y}$, defined as $y_{e'_i}=y_{e''_i}=x_{e_i}$ for every $i\in [m]$, satisfies $\mathbf{y} \in \Pea(G')$. Indeed, for every $i \in [n]$ we have that $\sum_{v'_i\ni e}y_e =\sum_{v''_i\ni e}y_e= \sum_{v_i\ni e}x_e \le b_{v_i}$.

    Since $\mathbf{y} \in \Pea(G')$, and $G'$ is bipartite, by Lemma~\ref{lem:bipartiteintegral}, $\mathbf{y}$ is a convex combination of some
    $\mathbf{y}^{(1)}, \dots, \mathbf{y}^{(\ell)}$ such that for each $1 \le j \le \ell$ it holds that $\mathbf{y}^{(j)} \in \Pea(G')$ and all entries of $\mathbf{y}^{(j)}$ are in $\{0,1\}$. Let $\lambda_1, \dots, \lambda_\ell \in [0,1]$ such that $y=\sum_{j\in [\ell]}\lambda_j \mathbf{y}^{(j)}$, and define for every $j \in [\ell]$, $\mathbf{x}^{(j)}$ such that $x^{(j)}_{e_i}=\frac 1 2 (y^{(j)}_{e'_i}+y^{(j)}_{e''_i})$ for every $i \in [m]$, which is half-integer. Note that for every $i \in [m]$, $x_{e_i}=\frac 1 2 (y_{e'_i}+y_{e''_i})=\frac 1 2 \sum_{j\in [\ell]}\lambda_j(y^{(j)}_{e'_i}+y^{(j)}_{e''_i})=\sum_{j\in [\ell]}\lambda_jx^{(j)}_{e_i}$ and, hence, $\mathbf{x}=\sum_{j\in [\ell]}\lambda_j\mathbf{x}^{(j)}$.
    Thus, $\mathbf{x}$ is a convex combination of $\mathbf{x}^{(1)}, \dots, \mathbf{x}^{(\ell)}$.
    We are left with showing that $\mathbf{x}^{(j)}$ belongs to $\Pea(G)$ for every $1 \le j \le \ell$.
    To do so note that for every $i \in [n], j \in [\ell]$,
    \[
    \sum_{v_i\ni e}x_e^{(j)} =\frac 1 2\sum_{v_i\ni e}y_{e'}^{(j)}+y_{e''}^{(j)}=\frac 1 2\left(\sum_{v'_i\ni e'}y_{e'}^{(j)}+\sum_{v''_i\ni e''}y_{e''}^{(j)}\right) \le b_{v_i},
    \]
which implies that $\mathbf{x}^{(j)} \in \Pea(G)$. Since $\mathbf{x}$ is a vertex of $\Pea(G)$ and a convex combination of $\mathbf{x}^{(1)}, \dots, \mathbf{x}^{(\ell)}$, it must be equal to one of them, say $\mathbf{x}=\mathbf{x}^{(j)}$ for some $j\in [\ell]$, and thus is half-integral.   
\end{proof}

\begin{proof}[Proof of Lemma~\ref{lem:noevencircuits}]
    Since every even cycle is an even circuit, it is straightforward to see that $G$ has no even cycle. Let us now assume that there exist two odd cycles $C_1$ and $C_2$ that share a node $u$ in $G$.
    Suppose that $C_1$ and $C_2$ share no edge.
    Then, the circuit that starts at $u$ and visits the first cycle, then the second, is an even circuit, a contradiction.
    Thus, $C_1$ and $C_2$ share at least one edge. Let $\{v,w\}$ be a shared edge such that the other neighbor of $w$ in $C_1$ is different from its neighbor in $C_2$ (this edge must exists otherwise $C_1=C_2$).
    Consider the path in $C_2$ from $w$ to $v$ that does not go through $\{v,w\}$. This path ends in $C_1$.
    Let $v'$ denote the first node on that path that is different from $w$ and is in $C_1$. We now have three edge-disjoint paths from $w$ to $v'$: two in $C_1$ and one in $C_2$. Two of them must have the same parity and thus together form an even circuit.
\end{proof}

\begin{proof}[Proof of Lemma~\ref{lem:sparsification1}]
Wajc~\cite{DBLP:conf/stoc/Wajc20} has proven that choosing edges using that method (that is, choosing colors uniformly at random then picking all edges of those colors in a proper coloring) ensures that any two adjacent edges are negatively associated, and that for any $v\in V, e' \ni v$, the random variables $\{[X_e|X_{e'}] \mid e \ni v\}$ are negatively associated.
Let $e$ be an edge of $E_i$. We have three cases:

\textbf{Case 1}: If $d\ge (1+\epsilon)^{i-1}$, then all colors are sampled and $H_i=E_i$. Moreover, $x_ed\ge (1+\epsilon)^{-i} (1+\epsilon)^{i-1} = \frac 1{1+\epsilon}$, and (\ref{eq:probainH}) holds trivially.

\textbf{Case 2}: If $x_e > \frac 1 d$, then $(1+\epsilon)^{-i+1} \ge x_e>\frac 1 d$ and in particular $d \ge (1+\epsilon)^{i-1}$, we thus refer to the previous case.

\textbf{Case 3}: If $x_e \le \frac 1 d$ and $d \le (1+\epsilon)^{i-1}$, then  $e$ is sampled with probability $\Proba[e \in H_i] =\frac {3\ceil{kd}}{3\ceil{k(1+\epsilon)^i}}$.
Since $kd \ge \frac 1 {\epsilon}$, we have that $kd+1 \le kd(1+\epsilon)$, and thus:
\[
\Proba[e \in H_i] =\frac {3\ceil{kd}}{3\ceil{k(1+\epsilon)^i}} \le \frac {kd+1} {k(1+\epsilon)^i} \le \frac {kd(1+\epsilon)} {k(1+\epsilon)^i} \le x_e \cdot d \cdot (1+\epsilon) = \min\{1, x_e\cdot d\} \cdot (1+\epsilon)
\]
On the other hand, since $(1+\epsilon)^i\ge (1+\epsilon)^{i-1}\ge d \ge \frac 1 {k\epsilon}$, we have that $k(1+\epsilon)^i +1\le k(1+\epsilon) ^{i+1} $. Therefore,
\[
\Proba[e \in H_i] =\frac {3\ceil{kd}}{3\ceil{k(1+\epsilon)^i}} \ge \frac {kd} {k(1+\epsilon)^i+1} \ge \frac {kd} {k(1+\epsilon)^{i+1}} \ge \frac {x_e \cdot d} { (1+\epsilon)^2} = \frac {\min\{1, x_e\cdot d\}} { (1+\epsilon)^2}.
\qedhere
\]
\end{proof}

For the proof of Lemma~\ref{lem:sparsification}, we need the following inequality:
\begin{theorem}[Bernstein's Inequality for Negatively Associated Variables]\label{thm:bernstein}
Let $Y$ be the sum of negatively associated random variables $Y_1, \dots, Y_\ell$, with $Y_i \in [0, U]$ for each $i \in [\ell]$. Then, for $\sigma^2=\sum_{i=1}^\ell \Var (Y_i)$ and all $a>0$,
\[
\Proba[Y>\expect[Y]+a]\le \exp\left(\frac{-a^2}{2(\sigma^2+aU/3)}\right)
\]
\end{theorem}

\begin{proof}[Proof of Lemma~\ref{lem:sparsification}]
Let $\mathbf{z} \in \mathbb{R}^E$ with $z_e :=\frac {x_e(1-3\epsilon)}{\min\{1, x_e \cdot d\}}\cdot X_e$.
By equation (\ref{eq:probestimate}),
\begin{equation}\label{eq:zlarge}
   \expect[z_e] = \expect [z_e | X_e] \cdot \Proba [X_e] \ge x_e(1-3\epsilon)\cdot (1+\epsilon)^{-2}\ge x_e (1-5\epsilon) 
\end{equation}
Therefore, $\mathbf{z}$ is a good approximation for $\mathbf{x}$ in the sense that $\expect\left[\sum_e z_e\right]\ge \sum_e x_e(1-5\epsilon)$. However, as we are scaling up $x_e$ to get $z_e$ for some edges $e$, $\mathbf{z}$ might not be a feasible fractional $k$-matching.
We obtain a feasible fractional $k$-matching $\mathbf{y}$ from $\mathbf{z}$
as follows:
\[
y_e:=\left\{\begin{array}{ll}
    0 & \text{if } x_e < 1/d \text{ and } \max_{v\in e}\left\{\sum_{e'\ni v}z_{e'}\right\}>k \\
    z_e & \text{otherwise}
\end{array}\right.
\]

\begin{claim}
    $\mathbf{y}$ is a feasible fractional $k$-matching.
\end{claim}

\begin{claimproof}
    Consider a node $v$. If $\sum_{e\ni v}z_{e} \le k$, then $\sum_{e\ni v}y_{e}\le \sum_{e\ni v}z_{e} \le k$.
    Otherwise, if $\sum_{e\ni v}z_{e}>k$, then $\sum_{e\ni v}y_{e} = \sum_{e\ni v, x_e \ge 1/d}z_{e} = \sum_{e\ni v, x_e \ge 1/d}x_e(1-3\epsilon)\le k$.
\end{claimproof}

To complete the proof, we will now show that for every edge $e$, we have $\expect[y_e] \ge (1-\epsilon) \expect [z_e]$.
If $x_e\ge \frac{1}{d}$, this trivially follows since $y_e=z_e$ and thus $\expect[y_e] = \expect [z_e]$.
We thus concentrate on the case $x_e < \frac{1}{d}$ and bound by $(1-\epsilon)$ the probability of the event $y_e \neq z_e$, that is the event $\max_{v\in e}(\sum_{e'\ni v}z_{e'})>k$.
In particular, we will consider the case $X_e=1$.

Let $v$ be an endpoint of $e$. We have that $x_e < 1/d \le k\epsilon$ (because we choose $d$ such that $d \ge 1/k\epsilon$). Let $e'\neq e$ such that $v \in e'$. Since $X_e$ and $X_{e'}$ are negatively correlated, we have that $\Proba[X_{e'}|X_e] \le \Proba [X_{e'}]\le \min\{1, x_{e'}\cdot d\}\cdot(1+\epsilon)$, by equation (\ref{eq:probestimate}).
Therefore:
\[\expect[z_{e'}|X_e] = \frac{x_{e'}(1-3\epsilon)}{\min\{1, x_{e'}d\}}\cdot \Proba[X_{e'}|X_e]\le \frac{x_{e'}(1-3\epsilon)}{\min\{1, x_{e'}d\}}\cdot \Proba[X_{e'}] \le x_{e'}(1-3\epsilon)(1+\epsilon) \le x_{e'}(1-2\epsilon)\]

Hence:
\[
\expect\left[\sum_{e' \ni v}z_{e'} \Big| X_e\right]=\expect[z_e |X_e] + \sum_{e' \ni v, e \neq e'}\expect[z_{e'}|X_e]
\le k\epsilon + \sum_{e' \ni v, e \neq e'}x_{e'}(1-2 \epsilon) \le k(1-\epsilon)
\]
We therefore expect $\mathbf{z}$ to not violate the constraint on $v$.
To bound the probability that $\mathbf{z}$ does violate the constraint, we first compute the variance of $\left[\sum_{e' \ni v}z_{e'}|X_e\right]$, and in particular, for every $e' \ni v$, the variance of $[z_{e'}|X_e]$.

If $e'$ is such that $x_{e'}\ge \frac{1}{d}$, then $\Proba [X_{e'}]=1$, and in particular $\Proba [X_{e'}|X_e]=1$. The variance of $z_e=\frac {x_e(1-3\epsilon)}{\min\{1, x_ed\}}\cdot X_e$ is therefore $0$. On the other hand, if $x_{e'} < 1/d$, then $[z_{e'}|X_e]$ is a Bernoulli random variable scaled by $\frac{1-3\epsilon}{d}$, with success probability at most $\Proba [X_{e'} \big| X_e]\le min\{1,x_{e'}d\}\cdot(1+\epsilon)=x_{e'}d(1+\epsilon)$. Therefore, the variance of this random variable is at most
\[
\Var([z_{e'}|X_e])\le \left(\frac {1-3\epsilon} d\right) ^2 \cdot x_{e'}d(1+\epsilon)\le \frac {x_{e'}} d.
\]

Summing over all edges, we get:
\[
\sum_{e'\ni v}\Var\left([z_{e'}|X_e]\right)\le \sum_{e'\ni v}\frac {x_{e'}} d \le \frac k d
\]

Since the variables $\{[X_{e'}|X_e] \mid e'\ni v\}$ are negatively associated, so are the variables $\{[z_{e'} \mid X_e]|e'\ni v\}$, by closure of negative association under scaling by positive constants. Therefore, we can use Bernstein's inequality (cf.~Theorem~\ref{thm:bernstein}).

For $\left[\sum_{e' \ni v}z_{e'} \big| X_e\right]$ to go over $k$, it needs to exceed its expectation by at least $k\epsilon$. Since for all $e'$ such that $x_{e'} > 1/d, [z_{e'}|X_e]$ is deterministic, this is equivalent to the sum only over edges $e'$ such that $x_{e'} \le 1/d$ exceeding its expectation by at least $k\epsilon$. In that case, $z_{e'} \le\frac {(1-3\epsilon)} d \le \frac 1 d$.
We thus have
\begin{multline*}
    \Proba\left[\sum_{e' \ni v}z_{e'}>k \bigg| X_e\right]
    \le \Proba\left[\sum_{e' \ni v, x_{e'} \le \frac 1 d}z_{e'}>\expect\Big[\sum_{e' \ni v, x_{e'} \le \frac 1 d}z_{e'}\big|x_e\Big]+\epsilon k\Bigg|X_e\right]\\
    \le \exp\left(-\frac{\epsilon^2k^2}{2\cdot(k/d+\epsilon k/3d)}\right) \le \exp\left(-\frac{\epsilon^2k}{2\cdot(1/d+\epsilon/3d)}\right) \le \exp\left(-\frac{\epsilon^2k}{4d}\right),
\end{multline*}
which is at most $\epsilon/2$ since $d \ge \frac{4\log(2/\epsilon)}{k\epsilon^2}$.

For $y_e$, we thus have that, conditioned on $e \in H$, the probability of the constraints on each of the nodes of $e$ being violated is at most $\epsilon$.
By union bound, we thus have $\Proba[y_e = z_e | X_e]\ge (1-\epsilon)$. Combined with Equation (\ref{eq:zlarge}), we have:
\[
    \expect[y_e]= \frac{x_e(1-3\epsilon)}{\min\{1, x_ed\}}\cdot \Proba[y_e=z_e|X_e]\cdot\Proba[X_e]
    \ge (1-\epsilon) \cdot \expect [z_e] \ge x_e(1-6\epsilon)
\]

We thus conclude that $H$ contains a fractional $k$-matching of expected value at least $1-6 \epsilon$ times the value of the fractional $k$-matching $\mathbf{x}$ in $G$.
\end{proof}

\begin{proof}[Proof of Lemma~\ref{lem:amortize}]
Let $f^*$ be an optimal coloring before the updates, and $g^*$ be an optimal coloring after the updates.
Let
$q :=\card{g^{-1}([k])}$,
$p^* :=\card{{f^*}^{-1}([k])}$, and
$q^* :=\card{{g^*}^{-1}([k])}$.
Since every deleted edge decreases the %
number of colored edges by at most $1$, we have:
\[
q\ge p-\floor{\epsilon p} \ge p(1-\epsilon)
\]
Similarly, since every added edge increases the size of the optimal coloring by at most $1$, we have:
\[
q^*\le p^*+\floor{\epsilon p} \le p^*(1+\epsilon)
\]
If $f$ is an $\alpha$-approximation, then $p^* \le \alpha \cdot p$.
Therefore, as long as $\epsilon \le \frac 1 3$:
\[
q^* \le p^*(1+\epsilon) \le \alpha \cdot p(1+\epsilon) \le \alpha \cdot \frac {1+\epsilon}{1-\epsilon} \cdot q \le \alpha \cdot q (1+3\epsilon)
\]
Hence, whenever we compute a partial coloring of a dynamic graph of size $p$,
we can charge the cost of that computation to the next $\floor{\epsilon p}$ updates without recomputing anything, and while losing a $(1+3\epsilon)$ approximation factor at most.
\end{proof}

%% file: 22-related-work.tex
\section{Related Work}\label{sect:related}
In this section, we give a more extensive overview over related work.
As the maximum $k$-edge coloring problem is also closely related to various matching problems,
we include relevant results for these problems. %
\subparagraph*{Edge Coloring}
Given a graph $G$, its \emph{chromatic index} $\chi'(G)$ is the smallest value $q$
such that all edges of $G$ can be colored with $q$ colors.
It is straightforward that $\Delta(G) \leq \chi'(G)$, where $\Delta(G)$ denotes the
maximum vertex degree in $G$.
Vizing~\cite{vizing1964} showed that $\chi'(G) \leq \Delta(G)+1$.
For bipartite graphs, $\chi'(G) = \Delta(G)$~\cite{konig1916graphen}.
In general, it is NP-hard to decide whether a given graph $G$ has $\chi'(G) = \Delta(G)$
or $\chi'(G) = \Delta(G) + 1$ already for $\Delta(G) = 3$~\cite{DBLP:journals/siamcomp/Holyer81a},
and even if $G$ is regular~\cite{nphard}.
Note that if $\Delta(G) = 1$, then $G$'s edges form a matching, whereas if $\Delta(G) = 2$,
then $G$ is a collection of cycles and paths and $\chi'(G) = 2$ iff all cycles have even length.
Another lower bound on the chromatic index is given by the \emph{odd density} $\rho(G)
:= \max_{S \subseteq V, |S| = 2i+1} \left\lceil\frac{E(S)}{i}\right\rceil$, where $E(S)$ denotes the
set of edges in the subgraph induced by $S$:
As all edges of the same color form a matching, at most $\left\lfloor \frac{|S|}{2}\right\rfloor = i$
edges of $E(S)$ can share the same color, so at least $\left\lceil\frac{E(S)}{i}\right\rceil$ colors are necessary
to color $E(S)$.
Edmonds~\cite{matching-theory} showed that the \emph{fractional chromatic index} is equal to $\max\{\Delta(G), \rho(G)\}$.

Misra and Gries~\cite{misragries} designed an algorithm that uses at most
$\Delta(G)+1$ colors.
It processes the edges in arbitrary order and colors each in $O(n)$ time, thus
resulting in an $O(mn)$ running time overall.
If new edges are added to the graph, they can be colored in $O(n)$ time.
Their algorithm improved on an earlier approach by
Gabow~\cite{gabowcolor}, which has a running time of $O(m \Delta(G) \log n)$.
Simmanon~\cite{DBLP:journals/corr/abs-1907-03201} reduces the time for finding
a $(\Delta(G)+1)$-edge coloring to $O(m\sqrt{n})$. By allowing more colors -- up to $(1+\epsilon)\Delta$ colors -- Duan, He and Zhang~\cite{DBLP:conf/soda/DuanHZ19} further reduce the running time to $O(m\cdot \poly(\log n, \epsilon^{-1}))$ as long as $\Delta(G)= \Omega(\log^2n\cdot \epsilon^{-2})$.
For bipartite graphs, Cole, Ost, and Schirra~\cite{bipartitecolouring} gave
an optimal algorithm (it uses $\Delta$ colors) with $O(m \log \Delta(G))$ running time.

Cohen, Peng, and Wajc~\cite{DBLP:conf/focs/CohenPW19} recently studied the
edge coloring problem in the online setting and proved various competitive
ratio results.

For the dynamic setting, Bhattacharya, Chakrabarty, Henzinger, and Naongkai~\cite{deltaedgecolouring}
show how to maintain a $(2\Delta(G)-1)$-edge coloring in $O(\log n)$ worst-case update time.
They also show that a $(2+\epsilon)\Delta(G)$-edge coloring can easily be
maintained with $O(1/\epsilon)$ expected update time. If $\Delta(G)= \Omega(\log^2n\cdot \epsilon^{-2})$, Duan, He and Zhang~\cite{DBLP:conf/soda/DuanHZ19} maintain an edge-coloring using $(1+\epsilon)\Delta$ colors in amortized $O(\log^8n\cdot \epsilon^{-4})$ update time.

\subparagraph*{Maximum $k$-Edge Coloring}
For the maximum $k$-edge coloring problem, the number of available colors is limited to
some $k \in \mathbb{N}$ and the task is to find a maximum-cardinality subset of
the edges $H$ such that for the subgraph restricted to $H$, $\Restrict{G}{H}$,
$\chi'(\Restrict{G}{H}) \leq k$.

This problem was first studied
by Favrholdt and Nielsen~\cite{DBLP:journals/algorithmica/FavrholdtN03}
in the online setting.
They propose and analyze the competitive ratio of various online algorithms and
show that every algorithm that never chooses to not color (``reject'') a colorable edge
has a competitive ratio between $0.4641$ and $\frac{1}{2}$,
and that any online algorithm is at most
$\frac{4}{7}$-competitive.

Feige, Ofek, and Wieder~\cite{originalpaper} considered the $k$-edge coloring
problem in the static setting and for multigraphs, motivated by a call-scheduling problem in
satellite-based telecommunication networks.
The authors show that for every $k \geq 2$, there exists an $\epsilon_k > 0$ such that
it is NP-hard to approximate the problem within a ratio better than $(1+\epsilon_k)$.
They also describe a static $(1-(1-1/k)^k)^{-1}$-approximation algorithm for general $k$
as well as a $\frac{13}{10}$-approximation for $k=2$.
The former algorithm applies a greedy strategy and works by repeatedly
computing a maximum-cardinality matching $M$, then removing $M$ from graph.
As all edges in a matching can be colored with the same color, $k$ repetitions
yield a $k$-edge coloring.
They also note that for a multigraph of multiplicity $d$, a
$\frac{k+d}{k}$-approximate solution can be obtained by first computing a $k$-matching,
then coloring the subgraph using $k+d$ colors (which is always possible, in analogy to
Vizing's theorem), and then discarding the $d$ colors that
color the fewest edges.
For simple graphs (i.e., $d=1$), this yields an approximation ratio of $\frac{k+1}{k}$.
The authors also give an algorithm with an approximation ratio tending to $\frac{1}{\alpha}$
as $k \rightarrow \infty$, where $\alpha$ denotes the best approximation ratio for
the chromatic index in multigraphs.

Several improved approximation results for the cases where $k=2$ and $k=3$ exist.
The currently best ratios are $\frac{6}{5}$ for $k=2$ and $\frac{5}{4}$ for $k=3$~\cite{koverkplus1}.

The maximum $k$-edge coloring problem was first studied in the edge-weighted setting by Hanauer,
Henzinger, Schmid, und Trummer~\cite{infocomstatic}.
Here, instead of finding a maximum-cardinality subset of the edges, the total weight
of the colored edges is to be maximized.
The authors describe several approximation and heuristic approaches to tackle
the problem in practice and provide an extensive experimental performance
evaluation on real-world graphs.
They also show that a double-greedy approach, where successively $k$ weighted matchings are computed by
a greedy algorithm, yields a $O(1)$-approximation. 
In a follow-up work, Hanauer, Henzinger, Ost, and Schmid~\cite{infocomdynamic}
design a collection of different dynamic and batch-dynamic algorithms for the weighted $k$-edge coloring.
Their focus is again more on the practical side.
Ferdous~\etal{}~\cite{DBLP:conf/esa/FerdousSPHK24} recently studied the problem in the streaming setting.

\subparagraph*{Matching}
The matching problem in the static setting has been subject to extensive
research both in the unweighted and weighted case.
The currently fastest deterministic algorithms for unweighted matching in
general and bipartite graphs have a running time of $O(m\sqrt{n})$~\cite{DBLP:conf/focs/MicaliV80,DBLP:journals/jacm/GabowT91,DBLP:journals/siamcomp/HopcroftK73}.
For weighted matching on general graphs, the currently best running time is
$O(n(m + n\log n))$~\cite{DBLP:conf/soda/Gabow90}.
An excellent overview, which also encloses approximation ratios, is given by
Duan and Pettie~\cite{DBLP:conf/focs/DuanP10}.

For the dynamic setting, Onak and Rubinfeld~\cite{DBLP:conf/stoc/OnakR10}
present a randomized $O(1)$-approximation algorithm with $O(\log^2 n)$ update
time.
The algorithm by Baswana, Gupta, and Sen~\cite{DBLP:conf/focs/BaswanaGS11}
improves the running time to $O(\log n)$ and the approximation ratio to $2$.
Later, Solomon~\cite{DBLP:conf/focs/Solomon16} reduced the amortized expected
update time to $O(1)$.
Wajc~\cite{DBLP:conf/stoc/Wajc20} gives a metatheorem for rounding a dynamic
fractional matching against an adaptive adversary and a $(2 +
\epsilon)$-approximate algorithm with constant update time or $O(\poly(\log n,
\epsilon^-1))$ worst-case update time.

Neiman and Solomon~\cite{DBLP:conf/stoc/NeimanS13} show that a $\frac{3}{2}$-approximate
matching can be maintained deterministically in $O(\sqrt{m})$ worst-case update time.
Bhattacharya, Henzinger, and Italiano~\cite{DBLP:conf/soda/BhattacharyaHI15}
give a deterministic $(3+\epsilon)$-approximation with $O(m^{1/3}\epsilon^{-2})$
amortized update time, as well as an $(4+\epsilon)$-approximation with
$O(m^{1/3}\epsilon^{-2})$ worst-case update time.
An improved algorithm is given by Bhattacharya, Henzinger, and
Nanongkai~\cite{DBLP:conf/stoc/BhattacharyaHN16}, which has $O(\poly(\log n,
\frac{1}{\epsilon}))$ amortized update time and an approximation ratio of
$(2+\epsilon)$.
A $(1+\epsilon)$-approximation algorithm with $O(\sqrt{m}\epsilon^{-2})$ worst-case update
time is given by Gupta and Peng~\cite{oneplusepsilonmatching} for $\epsilon < \frac{1}{2}$.
The authors also give an algorithm for the weighted case with the same approximation
ratio and a worst-case update time of
$O(\sqrt{m}\epsilon^{-2-O(1/\epsilon)}\log W)$, where $W$ is the largest weight
of an edge in the graph.

For bipartite graphs, Bosek, Leniowski, Sankowski, and
Zych~\cite{DBLP:conf/focs/BosekLSZ14} give a partially dynamic algorithm for
either the insertions-only or deletions-only setting, which runs in $O(m\sqrt{n})$
total time, thus matching the time of the Hopcroft-Karp static algorithm~\cite{DBLP:journals/siamcomp/HopcroftK73}. Due to the direct reduction of matching to maximum flow, the static case can now be solved in $O(m^{1+o(1)})$ time thanks to a breakthrough result of Chen, Kyng, Liu and Peng~\cite{DBLP:conf/focs/ChenKLPGS22}.

Stubbs and Williams~\cite{metatheoremsone} show how to transform a dynamic
$\alpha$-approximation algorithm for the unweighted matching problem to
a $(2+\epsilon)\alpha$-approximation algorithm for the weighted setting.
The running time increases by a factor of $\epsilon^{-2}\log^2 W$, where $W$
denotes the maximum weight of an edge.
Bernstein, Dudeja, and Langley~\cite{metatheoremtwo} improve on this result
w.r.t.\ running time and show that a $(1+\epsilon)\alpha$-approximation algorithm
can be obtained in the case of bipartite graphs.

Various results~\cite{DBLP:conf/focs/Behnezhad21} also exist for the ``value''
version of the problem, where one is only interested in the maximum size or
weight, but not in the set of edges.

\subparagraph*{$b$-Matching}
Gabow~\cite{gabowbmatch} gives a $O(\sqrt{\lVert b\rVert_1}m)$-time algorithm to compute a $\bvec{b}$-matching
in the unweighted, static setting.
If the $\bvec{b}$-matching is weighted, the running time is $O(\lVert b\rVert_1\cdot\min(m\log n, n^2))$.
Ahn and Guha~\cite{bmatchingapprox} give an algorithm that computes an $(1+\epsilon)$-approximation for $\bvec{b}$-matching and runs in $O(m \poly(\log n, \epsilon^{-1}))$ time.

For dynamic graphs, Bhattacharya, Henzinger, and Italiano~\cite{primaldual}
give a deterministic algorithm that maintains an $O(1)$-approximate fractional $k$-matching
with $O(\log^3 n)$ amortized update time.
This result is improved by Bhattacharya, Gupta, and
Mohan~\cite{BhattacharyaGM17}, who show how to maintain an integral
$(2+\epsilon)$-approximate $\bvec{b}$-matching in expected amortized $O(1/\epsilon^4)$
update time, against an oblivious adversary.